\documentclass[prl,preprint,endfloats]{revtex4}
\usepackage{graphicx}
\usepackage{bm}
\usepackage{pifont}
\usepackage{amsmath}

\begin{document}

\title{Structure analysis of high-pressure phase for skyrmion-hosting multiferroic Cu$_2$OSeO$_3$}

\author{E. Nishibori$^{1}$} 
\author{S. Karatsu$^{1}$} 
\author{C. Terakura$^{2}$}
\author{N. Takeshita$^{3}$}
\author{M. Kinoshita$^{4}$}
\author{S. Ishiwata$^5$}
\author{Y. Tokura$^{2,4,6}$} 
\author{S. Seki$^{2,4,7,8}$}

\affiliation{$^1$ Faculty of Pure and Applied Sciences and Tsukuba Research Center for Energy Materials Science (TREMS), University of Tsukuba, Tsukuba 305-8571, Japan}
\affiliation{$^2$ RIKEN Center for Emergent Matter Science (CEMS), Wako 351-0198, Japan}
\affiliation{$^3$ National Institute of Advanced Industrial Science and Technology (AIST), Tsukuba, Ibaraki 305-8562, Japan}
\affiliation{$^4$ Department of Applied Physics, University of Tokyo, Tokyo 113-8656, Japan}
\affiliation{$^5$ Division of Materials Physics, Graduate School of Engineering Science, Osaka University, Osaka, 560-8531, Japan}
\affiliation{$^6$ Tokyo College, University of Tokyo, Tokyo 113-8656, Japan}
\affiliation{$^7$ Institute of Engineering Innovation, University of Tokyo, Tokyo 113-8656, Japan}
\affiliation{$^8$ PRESTO, Japan Science and Technology Agency (JST), Kawaguchi 332-0012, Japan}

\begin{abstract}

Cu$_2$OSeO$_3$ is known as a unique example of insulating multiferroic compounds with skyrmion spin texture, which is characterized by the chiral cubic crystal structure at ambient pressure. Recently, it has been reported that this compound shows pressure-induced structural transition with large enhancement of magnetic ordering temperature. In the present study, we have investigated the detailed crystal structure in the high pressure phase, by combining the synchrotron X-ray diffraction experiment with the diamond anvil cell and the analysis based on the genetic algorithm. Our resuts suggest that the original pyrochlore Cu network is sustained even after the structural transition, while the orientation of SeO$_3$ molecule as well as the position of oxygen in the middle of Cu tetrahedra are significantly modified. The latter features may be the key for the reported enhancement of $T_c$ and associated stabilization of skyrmion phase at room temperature. 

\end{abstract}
%\pacs{75.30.Ds, 75.70.-i, 65.40.-b}
\maketitle

\begin{figure}
\begin{center}
\includegraphics*[width=14cm]{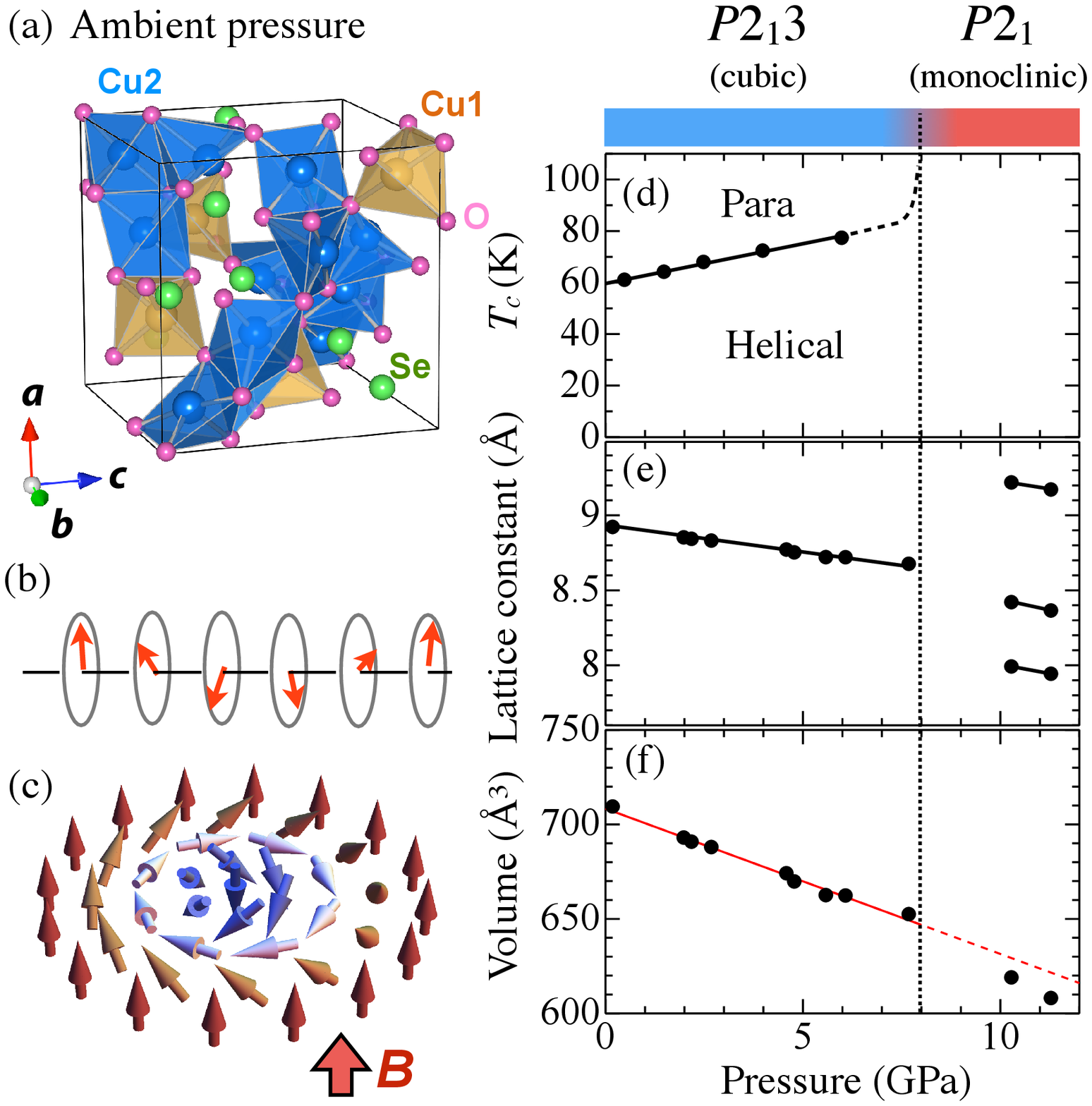}
\caption{(color online). (a) Crystal structure of Cu$_2$OSeO$_3$ at ambient pressure. (b) and (c) Schematic illustration of helical and skyrmion spin texture, respectively. (d)-(f) Pressure dependence of magnetic ordering temperature $T_c$, lattice constants and volume of crystallographic unit cell, respectively. In (f), the red solid line represents a linear fitting result for the data points in the $P2_1 3$ phase, which is extended into the $P2_1$ phase as the red dashed line. The data points in the latter phase show considerable deviation from this line, suggesting the further suppression of unit cell volume upon the structural phase transition. Note that the existence of intermediate phase with space group $P2_12_12_1$ has also been suggested for the narrow $P$-region between the $P2_13$ and $P2_1$ phases in Ref. \cite{COSO_PressurePNAS}, while this phase was not identified in the present experiments possibly due to the limited number of pressure data points.}
%\label{fig1}
\end{center}
\end{figure}

\begin{figure}
\begin{center}
\includegraphics*[width=10cm]{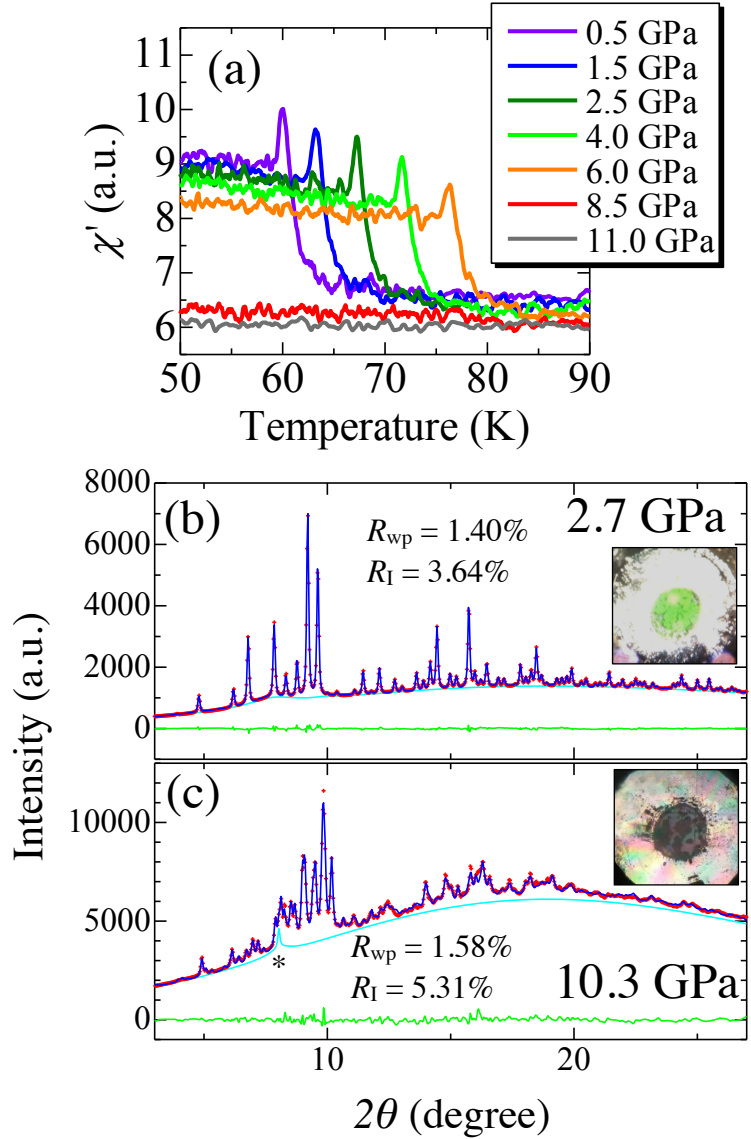}
\caption{(color online). (a) Temperature dependence of ac magnetic susceptibility $\chi '$, measured for the Cu$_2$OSeO$_3$ single crystal under zero magnetic field with various amplitude of hydrostatic pressure $P$. (b) and (c) Fitting results of Rietveld refinements for synchrotron X-ray powder diffraction data at (b) 2.7 GPa and (c)10.3 GPa, i.e. before and after the structural phase transition. Here, blue line, red + symbol, light blue line, and green line represent the experimental profile, the fitted profile, the background profile, and the deviation between the experimental and fitted ones, respectively. The * symbol indicates the peak from steal gasket. The measurement was performed at room temperature. The optical microscope image of the sample taken at each condition is also indicated. }
%\label{fig2}
\end{center}
\end{figure}

\begin{figure}
\begin{center}
\includegraphics*[width=11cm]{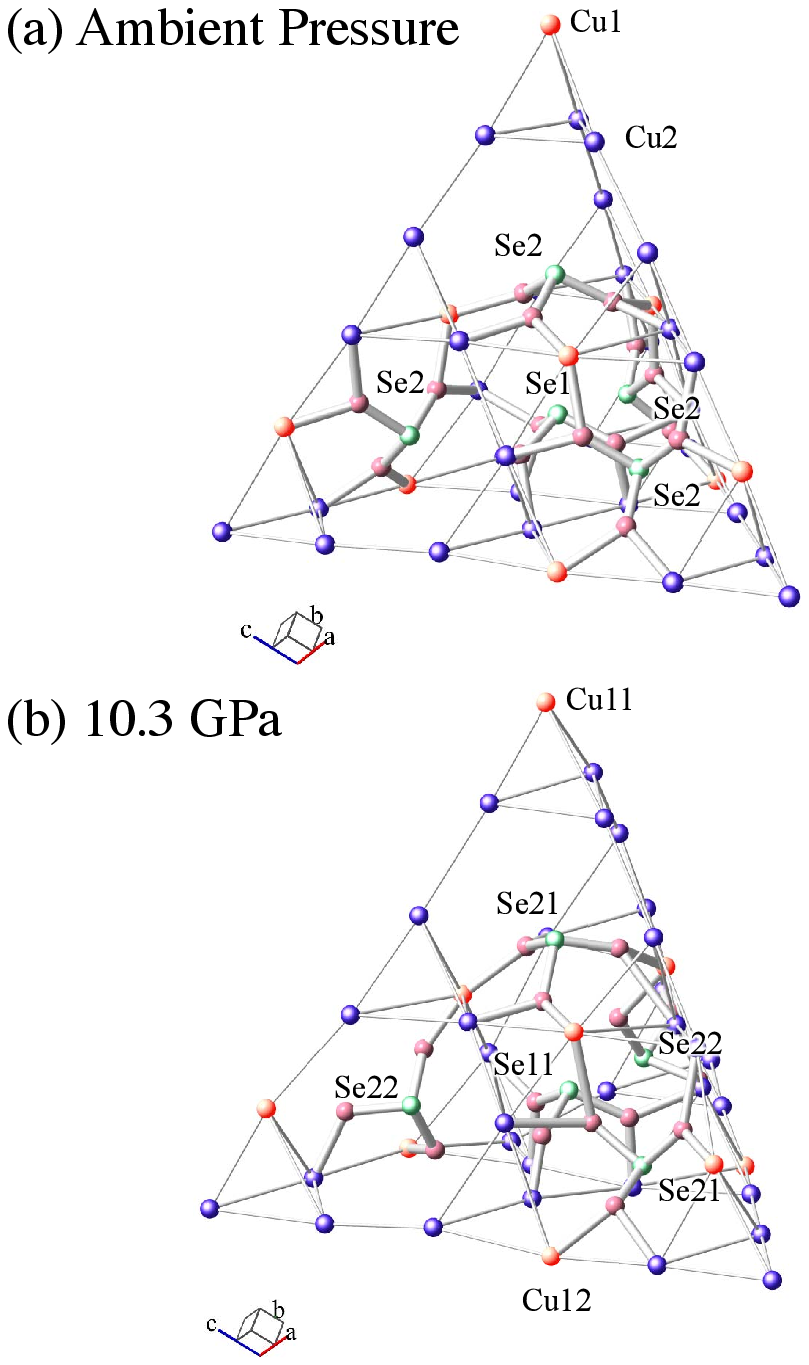}
\caption{(color online). The network of Cu ions and SeO$_3$ molecules at (a) ambient pressure and (b) 10.3 GPa (i.e. the high pressure phase). Other oxygen atoms are not shown here for clarity. Orange, blue, green and pink balls represent the Cu1, Cu2, Se and O atoms, respectively.}
%\label{fig3}
\end{center}
\end{figure}

\begin{figure}
\begin{center}
\includegraphics*[width=12cm]{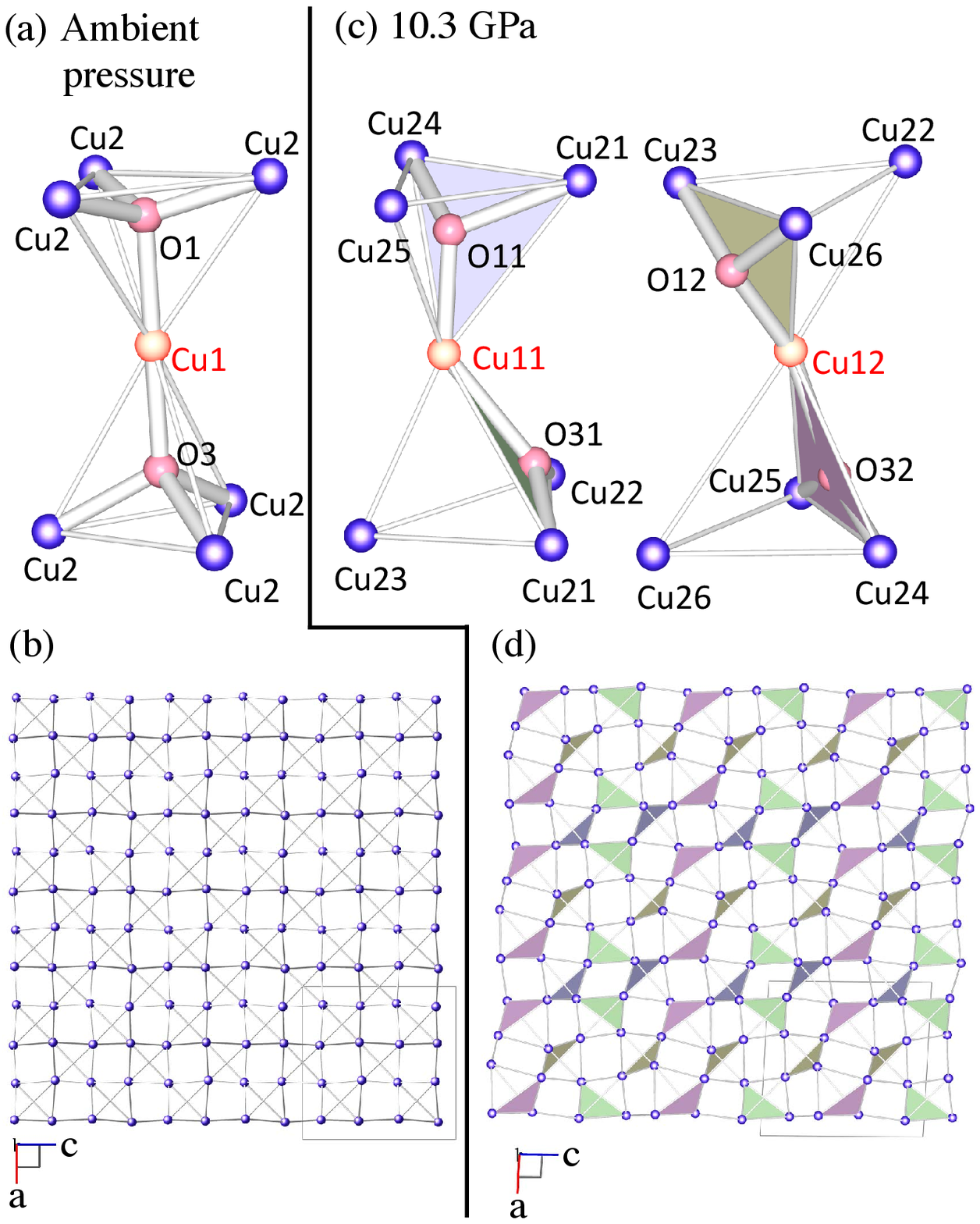}
\caption{(color online). (a) A pair of corner-shared Cu tetrahedra with the oxygen atoms located at their center and (b) the network of Cu ions viewed from the [010] direction, which are deduced from the structural analysis at ambient pressure. (c) and (d) The corresponding ones for $P=10.3$ GPa (i.e. the high pressure phase). The color of triangular plaquette in (d) is common with the one in (c).}
%\label{fig4}
\end{center}
\end{figure}

Magnetic skyrmion, i.e. a nanometric vortex-like swirling spin texture with topologically stable particle nature, has recently attracted much attention (Fig. 1(c))\cite{MnSi, TEMFeCoSi, SkXTheoryFirst, SkXReviewFertTwo, SkXReviewTokura}. In metallic systems, the interaction between conduction electrons and skyrmion spin texture enables the the unique manner of skyrmion manipulation by electric current\cite{CurrentControlNeutron, CurrentControlLTEM, CurrentControlBubble, THE, EmergentEfield}. Because of its particle-like character, small size and electric controllability, magnetic skyrmion is now intensively studied as a candidate of high-density information carrier for next generation of magnetic information storage and processing devices\cite{SkXReviewFertTwo, SkLogic, Neuromorphic}.

Previously, the observation of magnetic skyrmions has mostly been reported for the noncentrosymmetric systems, where Dzyaloshinskii-Moriya (DM) interaction plays a key role on the stabilization of skyrmion spin texture\cite{MnSi, TEMFeCoSi, SkXReviewTokura, Cu2OSeO3_Seki}. The DM interaction usually favors the helical magnetic order as the ground state (Fig. 1(b)), while the application of magnetic field $B$ induces the formation of the triangular lattice of skyrmions (Fig. 1(c)). The magnetic moment at the core (edge) of each skyrmion is aligned antiparallel (parallel) to external magnetic field, and the in-plane component of magnetic moments show various swirling spin texture depending on the symmetry of underlying crystal lattice\cite{MnSi, TEMFeCoSi, Cu2OSeO3_Seki, CoZnMn_First, GaV4S8, Heusler}.

So far, the most of skyrmion-hosting materials are reported to be metallic, while the electric-current-induced manipulation of skyrmions in such systems always causes the energy loss by the Joule heating. One potential solution is the employment of multiferroics, i.e. the insulating materials with both magnetic and dielectric orders. Our target compound Cu$_2$OSeO$_3$ is the first example of insulating materials to host magnetic skyrmions\cite{Cu2OSeO3_Seki}, which is characterized by the chiral cubic crystal structure with space group $P2_13$ at the ambient pressure (Fig. 1a)\cite{COSO_First, COSO_Structure}. In this compound, the skyrmion spin texture can magnetically induce electric polarization, through the symmetry reduction of charge distribution mediated by the spin-orbit interaction\cite{Cu2OSeO3_Seki, COSO_ME, COSO_Dielectric, COSO_DFT}. By utilizing such a strong coupling between the skyrmion spin texture and electric polarization, the electric-field-induced creation, annihilation\cite{COSO_Efield}, and motion\cite{COSO_Eresonance, COSO_Erotation} of skyrmions have successfully been demonstrated, which paves the way to the efficient skyrmion manipulation in insulators without Joule heat loss.

At this stage, the magnetic ordering temperature $T_c$ of such skyrmion-hosting multiferroics is relatively low ($T_c \sim 58$ K for Cu$_2$OSeO$_3$), and the further enhancement of their $T_c$ is highly demanded. Very recently, it has been reported that the application of hydrostatic pressure on Cu$_2$OSeO$_3$ leads to the structural phase transition with dramatic enhancement of $T_c$, where the stabilization of skyrmions at room temperature has been proposed\cite{COSO_PressurePNAS}. However, the detailed crystal structure in the high pressure phase, as well as the origin of the reported enhancement of $T_c$, remains yet to be resolved.

In this work, we studied the detailed crystal structure in the high pressure phase of Cu$_2$OSeO$_3$, through the synchrotron X-ray diffraction experiment with the diamond anvil cell and the associated analysis based on the genetic algorithm. Our results suggest that the original pyrochlore Cu network is sustained even after the structural transition, while the orientation of SeO$_3$ molecule as well as the position of oxygen in the middle of Cu tetrahedra are significantly modified. The latter features may be the key for the reported enhancement of $T_c$ and associated stabilization of skyrmion phase at room temperature.

The powder and single crystals of Cu$_2$OSeO$_3$ were prepared as described elsewhere\cite{COSO_InfraRed}. The high pressure X-ray powder diffraction profiles were measured with the diamond anvil cell. The culet size of the anvils was 1.0 mm and the diameter of stainless steel gasket hole was approximately 500 $\mu$m. The sample and several ruby small crystals with 2-3 $\mu$m size, whose fluorescence spectra allows the accurate evaluation of pressure amplitude, were loaded in the hole sealed with 4:1 methanol-ethanol solution as the pressure medium. This pressure medium can avoid the solidification at least up to 11 GPa (in contrast with the silicon oil used in Ref. \cite{COSO_PressurePNAS} that gradually solidifies above 3 GPa and causes inhomogeneous pressure distribution)\cite{PressureMedium}, which suppresses the broadening of diffraction peaks and enables the detailed structural analysis. The ac magnetic susceptibility was measured with the cubic-anvil-type hydrostatic pressure cryostat.

\begin{table}[h]
\caption{\label{table1} Atomic coordinates of Cu$_2$OSeO$_3$ at 10.3 GPa, which is characterized by the monoclinic space group $P2_1$ with lattice constants a=7.987(2)\AA, b=9.214(2)\AA, c=8.416(2)\AA and $\beta$=92.79(1)$^\circ$.
}
\begin{ruledtabular}
\begin{tabular}{lccccc}
$site$&$x$&$y$&$z$\\
\hline
Se11& -0.22072 & 0.71354 &  0.55357 \\
Se12&  0.24609 & 0.77457 & -0.00505 \\
Se21&  0.51611 & 0.94958 &  0.25412 \\
Se22&  0.02207 & 0.55026 &  0.20258 \\
Cu11& -0.12454 & 0.85879 &  0.17283 \\
Cu12&  0.36539 & 0.63550 &  0.36025 \\
Cu21& -0.10356 & 0.62461 &  0.88334 \\
Cu22& -0.33948 & 0.87528 &  0.88710 \\
Cu23&  0.61996 & 0.63833 &  0.13453 \\
Cu24&  0.13201 & 0.65734 &  0.64996 \\
Cu25& -0.14111 & 0.39966 &  0.59562 \\
Cu26& -0.44802 & 0.37754 &  0.32787 \\
O11 & -0.02562 & 1.00764 &  0.26310 \\
O12 &  0.55591 & 0.55314 &  0.32849 \\
O21 & -0.01885 & 0.74687 &  0.54087 \\
O22 & -0.23087 & 0.53680 &  0.52306 \\
O23 & -0.24531 & 0.72596 &  0.74661 \\
O24 &  0.38624 & 0.65268 &  0.05586 \\
O25 &  0.33637 & 0.92839 &  0.05198 \\
O26 &  0.26723 & 0.77650 & -0.19908 \\
O31 & -0.19316 & 0.79565 & -0.05429 \\
O32 &  0.17319 & 0.75362 &  0.46386 \\
O41 &  0.54470 & 0.77658 &  0.29676 \\
O42 &  0.52574 & 0.51164 &  0.56858 \\
O43 &  0.71269 & 1.00208 &  0.24616 \\
O44 &  0.01704 & 0.72671 &  0.17187 \\
O45 &  0.15378 & 0.50186 &  0.06769 \\
O46 &  0.14829 & 0.54426 &  0.36320 \\
\end{tabular}
\end{ruledtabular}
\end{table}

Figure 2(a) indicates the temperature dependence of ac magnetic susceptibility $\chi '$ for Cu$_2$OSeO$_3$, measured under zero magnetic field with various amplitude of hydrostatic pressure $P$. The $\chi '$ profile shows a clear peak structure at $T_c \sim 60$ K for 0.5 GPa, and the peak gradually shifts toward the higher temperature by increasing $P$. Above 8 GPa, the $\chi '$ peak structure suddenly disappears from the measured temperature range, which suggests the emergence of structural phase transition and associated large enhancement of $T_c$ as previously reported\cite{COSO_PressurePNAS, COSO_PressureRonnow, COSO_PressurePRB}. 

To investigate the pressure dependence of crystal structure, we have performed the synchrotron X-ray diffraction experiment at SPring-8 BL02B1 beamline with the diamond anvil cell. The wavelength of incident X-ray was 0.42734 \AA with 150 $\mu$m $\phi$ beamsize. Figures 2(b) and (c) indicate the powder diffraction profiles measured at 2.7 GPa and 10.3 GPa, respectively. The former profile is well reproduced by the structure at the ambient pressure (Fig. 2(b)). We observed the discontinuous change of powder diffraction pattern above 8 GPa as shown in Fig. 2(c). The latter high pressure phase is confirmed to be stable at least up to 11.3 GPa, which is the maximum pressure of the present experimental condition. Sharp distinct Bragg peaks in the high angle region of $2\theta>15^\circ$ (i.e. $d<1.5$ \AA) were recognized in Fig 2(c). These were crucial for following structural analysis.

Since the original crystal unit cell of Cu$_2$OSeO$_3$ at the ambient pressure (Fig. 1(a)) contains 56 atoms\cite{COSO_Structure}, the structural analysis based on the powder diffraction pattern under the high pressure is challenging. To identify the reasonable structural model to reproduce the observed diffraction pattern, we performed the analysis based on the genetic algorithm (GA) \cite{GA}, in which several basic molecular structural unit is assumed and their relative orientations and positions are explored as the fitting parameter. This approach has been particularly successful in the structural analysis of organic compounds, and the present Cu$_2$OSeO$_3$ containing the rigid SeO$_3$ molecule with $sp^3$ orbital can be an appropriate target. There are independent four SeO$_3$ molecules, eight Cu atoms and four oxygen atoms in the crystallographic asymmetric unit.  The number of degrees of freedom in one SeO$_3$ molecule is 6, including three positional parameters and three orientation parameters. Each isolated atom is characterized by three positional parameters. Therefore, the total number of degrees of freedom became 59, because one positoinal parameter must be fixed in the case of a noncentrosymmetric space group. After the search of the candidate structure by the genetic algorithm, the additional rigid-body Rietveld analysis\cite{Rietveld} with bond length and angle restraints has been performed to refine the structural parameter. Finally, we obtained the atomic coordinates with the reliability factor $R_{\rm wp} = 1.58 \%$ and $R_{\rm I} = 5.31 \%$ as summarized in Table 1 with the monoclinic space group $P2_1$. Similar Rietveld analysis has also been performed for various $P$, and we obtained the pressure dependence of lattice constants and volume of crystallographic unit cell as shown in Figs. 1(e) and (f). As the pressure increases, the unit cell volume linearly decreases and then shows a discontinuous drop at the structural phase transition. During this process, the number of atoms in the unit cell remains unchanged.

Figures 3(a) and (b) indicate the position of Cu$^{2+}$ ($S=1/2$) magnetic ions and SeO$_3$ molecules before and after the structural phase transition, respectively (Some oxygen atoms are not shown for clarity. The Cu frameworks viewed from the [010] direction are also indicated in Figs. 4(b) and (d)). In the original Cu$_2$OSeO$_3$ structure, Cu sites form a distorted pyrochlore lattice, which consists of the corner-shared Cu tetrahedra and large vacancy among them. We found that this Cu-pyrochlore framework is retained even in the high pressure phase, while its buckling manner is considerably modified (Figs. 4(b) and (d)). Here, the vacancy space is filled by the SeO$_3$ molecule. The shrinkage of vacancy volume leads to the tilting of SeO$_3$ orientation, which breaks the three-fold rotational symmetry in the system. 

In the following, we discuss the origin of magnetic order and its pressure-induced modification in Cu$_2$OSeO$_3$. The original Cu$_2$OSeO$_3$ structure contains two distinctive Cu$^{2+}$ sites surrounded by either a trigonal bipyramid or square pyramid of oxygen ligands (i.e. Cu1 and Cu2 sites, respectively) with the ratio of 1:3 (Fig. 1(a)). Here, the super-exchange interaction between the Cu1 and Cu2 is antiferromagnetic but the one between Cu2 ions is ferromagnetic, which stabilizes the three-up one-down type local ferrimagnetic order\cite{COSO_Dielectric, COSO_Ferri}. According to the previous calculation based on the density functional theory (DFT), a hole exists in the $d_{z^2}$ and $d_{x^2-y^2}$ orbitals for the Cu1 and Cu2 ions, respectively, and the sign of super-exchange interactions can be explained by the Kanamori-Goodenough rule with the Cu-O-Cu bonding angle close to 90 degree\cite{COSO_DFT}. Here, the most important change upon the structural transition is the shift of the oxygen atom position, that was originally located at the center of Cu-tetrahedra (Fig. 4(a)). In the high pressure phase, one of the four Cu-O bonds is broken and the oxygen moves to the middle of a triangular Cu plaquette as shown in Fig. 4(c). Such an oxygen position shift within the Cu tetrahedron, as well as the tilting of SeO$_3$ molecule whose oxygen atoms also contribute to the Cu-O-Cu super-exchange path (Fig. 3(b)), modifies the manner of crystal field splitting and associated orbital hybridization for each Cu-O-Cu bond, which would result in the dramatic change of magnetic interactions. Note that the original green transparent color of the sample (mainly reflecting the absorption by crystal field excitation\cite{COSO_Faraday}) turns into the black opaque upon the structural transition (Figs. 2(b) and (c)), which also supports the modification of orbital character in Cu sites. Because of the complicated crystal structure in the high pressure phase, the quantitative evaluation of exchange amplitude $J$ for each path is not straightforward. In general, the largest magnitude of $J$ is obtained when transfer integral for the super-exchange path is maximized, and we naively expect that such a situation probably emerges here. To fully clarify the origin of reported enhancement of $T_c$ and skyrmion formation at room temprature, further DFT calculation based on the proposed crystal structure would be helpful. 

In summary, we investigated the detailed crystal structure for the high pressure phase of skyrmion-hosting multiferroic Cu$_2$OSeO$_3$. Our analysis suggests that the tilting of SeO$_3$ molecule, as well as the modification of oxygen position within the Cu tetrahedra, are the main features of the pressure-induced structural transition. The present results set a fundamental basis for better understanding of the magnetism in the high pressure phase, and may contribute to the further development of the general strategy to realize multiferroic skyrmions at room temperature.

The authors thank T. Arima for enlightening discussions and K. Sugimoto and H. Kasai for experimental and analytical help. This work was partly supported by Grants-In-Aid for Scientific Research (grant nos 18H03685, 20H00349, 19KK0132 and 20H04656) from JSPS, PRESTO (grant no. JPMJPR18L5) and CREST (grant no. JPMJCR1874) from JST, Asahi Glass Foundation and Murata Science Foundation. The synchrotron radiation experiments were performed with the approval of the Japan Synchrotron Radiation Research Institute (JASRI) (Proposal Nos. 2015A1573. 2016A0078).

\end{document}